\documentclass[10pt,epsfig]{article}
\usepackage{amssymb}
\usepackage{graphicx}
\usepackage{amsmath}

\input epsf.sty
\newtheorem{theorem}{Theorem}
\newtheorem{acknowledgement}[theorem]{Acknowledgement}

\input{tcilatex}
\makeatletter \@addtoreset{equation}{section}
\renewcommand{\theequation}{\thesection.\arabic{equation}}

\begin{document}

\title{%
\rightline{\mbox {\normalsize
{Lab/UFR-HEP0403/GNPHE/0403/IFTCSIC/0422}}} \textbf{\textbf{RG Cascades in
Hyperbolic Quiver Gauge Theories}}\\
}
\author{R. Ahl laamara$^{1,4}$, M. Ait Ben Haddou$^{1,2,4}$, A Belhaj$^{3,4}$%
, L.B Drissi$^{1,4}$, E.H Saidi$^{1,4}$\thanks{%
h-saidi@fsr.ac.ma} \\
%EndAName
{\small 1 Lab/UFR-Physique des Hautes Energies, Facult\'{e} des Sciences de
Rabat, Morocco.}\\
{\small 2- D\'{e}partement de Mathematique, Facult\'{e} des Sciences de Mekn%
\`{e}s, Morocco}\\
{\small 3- Instituto de F\'{\i}sica Te\'{o}rica, C-XVI,} {\small Universidad
Aut\'{o}noma de Madrid, E-28049-Madrid, Spain}\\
{\small 4-Groupement National de Physique des Hautes Energies, GNPHE; Siege
focal, Rabat, Morocco.}}
\maketitle

\begin{abstract}
In this paper, we provide a general classification of supersymmeric QFT$_{4}$%
s into three basic sets: ordinary, affine and indefinite classes. The last
class, which has not been enough explored in literature, is shown to share
most of properties of ordinary and affine super QFT$_{4}$s. This includes,
amongst others, its embedding in type II string on local Calabi-Yau
threefolds. We give realizations of these supersymmetric QFT$_{4}$s as
D-brane world volume gauge theories. A special interest is devoted to
hyperbolic subset for its peculiar features and for the role it plays in
type IIB background with non zero axion. We also study RG flows and duality
cascades in case of hyperbolic quiver theories. Comments regarding the full
indefinite sector are made.

\end{abstract}
  \newpage
\tableofcontents

 \thispagestyle{empty}
  \newpage \setcounter{page}{1} \newpage

\section{Introduction}

The study of 4D supersymmetric quiver ADE gauge theories and their affine
extensions has attracted a lot of attention in the last few years. In fact,
this interest is mainly due to their specific realization as D-brane world
volume gauge theories \cite{1} and also to the role they play in low energy
type II superstring compactifcations \cite{2}-\cite{5}. Supersymmetric
quiver gauge theories were at the basis in deriving Dijkgraaf-Vafa (DV)
theory \cite{6}-\cite{8} and generally in the study of large N duality and
geometric transitions \cite{9} in type II string compactification on
Calabi-Yau threefolds (CY3) with ADE geometries \cite{10}-\cite{12}.

One of the objectives of the present study is to extend results of 4D $%
\mathcal{N}$$=2$ and $\mathcal{N}$$=1$ ordinary and affine ADE quiver gauge
models to more general supersymmetric gauge theories. The new class we will
consider here, which we dub \textit{indefinite class}, concerns those 4D
super QFTs that are associated with the indefinite sector of simply laced
Lie algebras \cite{13}-\cite{15}. Like for affine ADE models with their
special properties such as super-conformal invariance and RG cascades \cite%
{16}-\cite{19}, these new quiver gauge theories exhibit as well very
remarkable features despite they are not conformally invariant \cite{20}.
They appear as low energy field theory limit of type II strings on CY3 with
\textit{indefinite geometries} and have D-brane realizations quite similar
to those realizing affine models. Throughout this study, we also complete
the picture on 4D supersymmetric quiver gauge theories that may be embedded
in type IIB string compactified on more general local CY3s. In particular,
we show that indefinite symmetries have remarkably a special hyperbolic
sub-invariance which:\newline
(1) serves as a laboratory to explore the indefinite class of quiver gauge
theories, \newline
(2) push a step further Fiol analysis \cite{27} extending Klebanov-Strassler
affine A$_{1}^{2}$ model \cite{17,18} and,\newline
(3) appears, surprisingly, as a symmetry in type IIB background with non
zero axion.\newline
The last property gives an explicit evidence for the role played by
hyperbolic invariance in type IIB string theory and opens an issue to look
for hidden symmetries in string theory. We also study field dualities in
hyperbolic model and show that, like in affine case, these theories have
duality cascades.

The presentation of this work is as follows. In section 2, we give general
features on indefinite quiver gauge theories; in particular on the special
hyperbolic subset; its roots system and hyperbolic Weyl group. We show that
these objects play a key role in the geometric engineering of hyperbolic
quiver QFT$_{4}$s and their duals. In section 3, we develop the general
setting of indefinite 4D quiver gauge theories embedded in type IIB strings
on Calabi-Yau threefolds. Particular interest is given to the hyperbolic
subset. In section 4, we study the RG flows for hyperbolic theories and
analyze there duality cascades. Section 5 is devoted to summarize our
findings and to make an outlook. Two appendices on indefinite Lie algebras
and hyperbolic ADE quiver gauge models are given in section 6.

\section{Indefinite quiver gauge theories}

In this section we want to extend results of four dimensional supersymmetric
ADE quiver gauge theories and their affine extensions to other
supersymmetric gauge models with more general symmetries. To begin, note
that as far as $\mathcal{N}=2$ supersymmetric QFT$_{4}$s are concerned, one
immediately realizes that affine quiver gauge theories are not the unique
possible generalization of ordinary ADE QFT$_{4}$s. There are infinitely
many others that may be constructed, these concern the class of indefinite
quiver gauge theories that interests us here. A more interesting thing in
this issue is that the mysterious indefinite set contains the\ subset of
hyperbolic gauge models which have the property of being closer to the
standard affine Kac-Moody (KM) models.

In what follows, we give a classification theorem of all possible 4D
supersymmetric $\Pi _{i}U\left( N_{i}\right) $ quiver gauge theories. This
theorem allows us to have a full picture on 4D super-quiver field models and
opens a window on the role of indefinite Lie symmetries in field and string
theories. We will mainly focus on 4D $\mathcal{N}=2$ quiver gauge models to
elaborate the results, but by switching on non linear adjoint matter self
interactions, one gets the corresponding $\mathcal{N}=1$ ones.

\subsection{Classification of Super QFT$_{4}$s}

In classifying 4D supersymmetric quiver gauge theories, we use two standard
results:

(\textbf{1}) the field theoretical correspondence between Dynkin diagrams $%
\mathcal{D}\left( g\right) $ of Lie algebras $g$ on one hand and matter and
gauge content in 4D supersymmetric quiver gauge theories on the other hand
\cite{10}. This correspondence is defined by associating to each diagram $%
\mathcal{D}\left( g\right) $ of a ( simply laced ) Lie algebra, a quiver SQFT%
$_{4}$,%
\begin{equation}
\mathcal{D}\left( g\right) \mathcal{\qquad \longleftrightarrow \qquad }\text{%
Super-QFT}_{4}.
\end{equation}%
This means that degrees of freedom in supersymmetric quiver QFT$_{4}$ are
completely encoded in the graph $\mathcal{D}\left( g\right) $. In $\mathcal{N%
}=1$ superspace language, the explicit rule is as follows: (\textbf{a}) The
i-th node of $\mathcal{D}\left( g\right) $ encodes a gauge multiplet $V_{i}$
and a chiral matter $\Phi _{i}$ superfield both transforming in the adjoint
representation of $U\left( N_{i}\right) $. (\textbf{b}) Links $\left\vert
K_{ij}^{q}\right\vert $ between i-th and j-th nodes of $\mathcal{D}\left(
g\right) $ give the number of bi-fundamental matter chiral multiplets $%
Q_{ij}\oplus Q_{ji}$ transforming in $\left( N_{i},\overline{N}_{j}\right) $
representation of $U\left( N_{i}\right) \times U\left( N_{j}\right) $.

(\textbf{2}) Classification of possible Dynkin diagrams $\mathcal{D}\left(
g\right) $ of Lie algebras. This classification is given by two fundamental\
Lie algebraic theorems: (\textbf{a}) Vinberg theorem \cite{21} asserting
that in general there are three basic subsets of Dynkin diagrams which, for
convenience, we denote them as,
\begin{equation}
\mathcal{D}^{+}\left( g\right) \text{;\qquad }\mathcal{D}^{0}\left( g\right)
\text{;\qquad }\mathcal{D}^{-}\left( g\right) .  \label{clas}
\end{equation}%
These are respectively associated with finite dimensional Lie algebras,
affine KM extension and indefinite generalization. (\textbf{b}) Cartan type
classifications \cite{22,23} of above Dynkin diagrams eq(\ref{clas}) encoded
in the following statement (i) The usual Cartan classification of Dynkin
diagrams $\mathcal{D}^{+}\left( g\right) $,
\begin{equation}
\mathcal{D}^{+}\left( A_{r}\right) ,\qquad \mathcal{D}^{+}\left(
D_{r}\right) ,\qquad \mathcal{D}^{+}\left( E_{6,7,8}\right) ,
\end{equation}%
where A$_{r}$, D$_{r}$ and E$_{r}$ are the usual rank $r$ simply laced
finite dimensional Lie algebras. (ii) KM classification of affine Lie
algebras with co-rank one graphs $\mathcal{D}^{0}\left( g\right) $,
\begin{equation}
\mathcal{D}^{0}\left( \widehat{A}_{r}\right) ,\qquad \mathcal{D}^{0}\left(
\widehat{D}_{r}\right) ,\qquad \mathcal{D}^{0}\left( \widehat{E}%
_{6,7,8}\right) ,
\end{equation}%
where $\widehat{A}$ stands for the affine extension of $A$ and so on. (iii)
For the classification of Dynkin diagrams $\mathcal{D}^{-}\left( g\right) $
of indefinite simply laced Lie algebras, there have been many attempts but
unfortunately with partial results only since a complete classification does
not exist yet. However, there is an interesting partial classification which
can be used to approach the indefinite sector. This partial classification
\cite{24,14}, which will be considered in the present study, deals with the
W. Li hyperbolic Dynkin diagrams denoted as $\mathcal{D}^{-}\left(
H_{r}^{s}\right) $ where r and s are integers. Results in this matter are
reported in appendix A, for a general analysis and other conventional
notations see \cite{24,25}.

\subsection{Theorem}

As far as Dynkin graphs of simply laced Lie algebras are concerned, there
are three kinds of four dimensional supersymmetric quiver gauge theories:

(\textbf{1}) The usual four dimensional supersymmetric ordinary ADE quiver
gauge theories. Their content in gauge and matter super multiplets are
encoded by the ordinary simply laced ADE Dynkin diagrams according to the
rule given above. These theories are not scale invariant, they appear as
type II strings low energy limits and have realization as gauge theories in
the world volume of partially wrapped D-branes.

(\textbf{2}) The well common supersymmetric affine $ADE$ quiver gauge
theories which can be viewed as the simplest extension of the previous ones.
They are obtained by adding more gauge and matter multiplets according to
the structure of affine $ADE$ Dynkin diagrams. These supersymmetric quantum
field models are specific gauge theories, they may be conformal and have a
remarkable realization as D3-brane world volume gauge theories.

(\textbf{3}) An indefinite sector of supersymmetric quiver QFT$_{4}$s. These
gauge theories have not been enough explored in the literature, they can be
thought of as given by the set of all remaining possible generalizations of
ordinary quiver gauge theories except affine KM extensions. Indefinite
quiver gauge theories include the hyperbolic quiver gauge models as a
special subset.

\subsection{Supersymmetric Action}

From the classification above, we deduce that given a Cartan matrix $%
K^{q}\left( g\right) $ and Dynkin diagram $\mathcal{D}^{q}\left( g\right) $,
with $q=1,0,-1$ referring respectively to ordinary simply laced Lie
algebras, affine KM extension and indefinite generalization, one can\
determine the appropriate number of propagating degrees of freedom of the
corresponding supersymmetric quiver QFT$_{4}$. The rule is as follows \cite%
{5}: To the i-th node of $\mathcal{D}^{q}$, one associates a gauge multiplet
$V_{i}$ and a chiral matter one $\Phi _{i}$ transforming in adjoint
representation of $U\left( N_{i}\right) $. For $\left\vert
K_{ij}^{q}\right\vert $ ($K_{ij}^{q}=0,-1$ for present case), links between
i-th and j-th nodes of $\mathcal{D}^{q}$, one associates bi-fundamental
matter $Q_{ij}$ transforming in $\left( N_{i},\overline{N}_{j}\right) $
representation of $U\left( N_{i}\right) \times U\left( N_{j}\right) $. With
these superfield degrees of freedom, we can write down the general
supersymmetric action describing their classical dynamics. In the $N=1$
formalism, the general superfield action $\mathcal{S}_{q}$ reads as:
\begin{eqnarray}
\mathcal{S}_{q} &=&\int d^{4}xd^{4}\theta \left[ \sum_{i,j}Tr\left(
Q_{ij}^{\ast }\left[ \exp \left( K_{ij}^{q}V_{j}\right) \right]
Q_{ji}\right) \right] +\sum_{i}\zeta _{i}\int d^{4}xd^{4}\theta Tr\left(
V_{i}\right)  \notag \\
&&+\left( \int d^{4}xd^{2}\theta \left[ \sum_{i,j}G_{\left[ ij\right]
}^{q}\lambda _{ij}Tr\left( Q_{ij}\Phi _{j}Q_{ji}\right) +\sum_{i}Tr\left[
\mathrm{W}\left( \Phi _{i}\right) \right] \right] +hc\right)  \label{act} \\
&&+\left( \int d^{4}xd^{2}\theta \left[ \sum_{i}Tr\left( \Phi _{i}^{\ast
}e^{K_{ii}^{q}adV_{i}}\Phi _{i}\right) \right] +\int d^{4}xd^{2}\theta \left[
\sum_{i}\frac{1}{g_{i}^{2}}Tr\left( W_{\alpha }^{i}W_{i}^{\alpha }\right) %
\right] +hc\right) ,  \notag
\end{eqnarray}%
where $W_{i}^{\alpha }$ stand for the spinor gauge field strength. The
constants $g_{i}$, $\zeta _{i}$, $\lambda _{ij}$ as well as those contained
in $\mathrm{W}\left( \Phi _{i}\right) $, which is a polynomial in adjoint
matter ($\mathrm{W}=\sum z_{i}\Phi _{i}+m_{i}\Phi _{i}^{2}...$), are the
usual bare coupling constants. Obviously $\mathcal{N}=2$ supersymmetry
requires linear adjoint matter superpotentials; i.e $\mathrm{W}=\sum
z_{i}\Phi _{i}$, otherwise it is broken down to $\mathcal{N}=1$. Note that
the $z_{i}$s should be compared with the FI coupling $\zeta _{i}$; they will
play a crucial role in next sections when we study Seiberg like field
duality. Note also that since Cartan matrices of simply laced Lie algebras
are symmetric, they can be expressed in terms of the $T_{ij}^{q}$ triangle
matrices ($i\geq j$) and their transpose $T_{ji}^{q}$\ as $%
K_{ij}^{q}=T_{ij}^{q}+T_{ji}^{q}$. These $T_{ij}^{q}$s are also used in
writing the $G_{\left[ ij\right] }^{q}$ antisymmetric matrices appearing in
above relation as $G_{[ij]}^{q}=T_{ij}^{q}-T_{ji}^{q}$. Finally, observe
that the above quiver gauge theory has an interpretation in type IIB string
theory as world volume theories on regular D3-branes and partially wrapped
D5 ones. We will turn to this representation later when we consider the
brane realization of hyperbolic gauge theories. From the action above, one
can determine the superfield eqs of motion. For the matter's case and up on
dropping kinetic terms non sensitive to complex deformations, we obtain the
following superfield equations of motion,
\begin{eqnarray}
Q_{ij}Q_{ji}-Q_{ji}Q_{ij} &=&W_{i}^{^{\prime }},  \notag \\
Q_{ij}\Phi _{j} &=&\Phi _{i}Q_{ij}.
\end{eqnarray}%
For linear superpotential $W$, these eqs describe, for each gauge group
factor, an $O(-2)\oplus O(0)$ Calabi-Yau threefold and the quiver gauge
theory has $\mathcal{N}=2$ supersymmetry. For generic polynoms in $\Phi _{i}$%
s however, $\mathcal{N}=2$ supersymmetry is explicitly broken down to $%
\mathcal{N}=1$. The new geometry of the Calabi-Yau threefolds associated
with each gauge group factor is now $O(-1)\oplus O(-1)\rightarrow P^{1}$.
Following DV theory, this manifold may undergo a geometric transition where
the blow down $S^{2}$ spheres are replaced by $S^{3}$ and the partially
wrapped D5-branes by three form fluxes. At this level nothing requires a
distinction between indefinite quiver gauge theories and the ordinary and
affine ones. This property let understand that basic results on the familiar
supersymmetric quiver gauge theories are also valid for the indefinite case.
Therefore like for ordinary and affine ADE quiver gauge models which are
known as QFT limits of type II string on CY3 with ordinary and elliptic ADE
singularities respectively, it is natural to think about the indefinite
quiver gauge theories defined above as also following from type II string
compactification on CY3 with indefinite singularities. The structure of
these indefinite singularities and the underlying singularity theory is
still an open question; but we fortunately dispose of some explicit examples
realizing CY3 manifolds with indefinite singular CY3s. For details on this
issue; we refer to \cite{15,16}.

\section{Hyperbolic Quiver Gauge models}

Hyperbolic quiver gauge models constitute a special subset of indefinite
quiver gauge theories that are based on hyperbolic Lie algebras. Some of
these hyperbolic models may be also viewed as simplest extension of the
usual supersymmetric affine ADE theories. Moreover, they can be also
obtained as low energy limit of type IIB string on CY3s with hyperbolic
singularities and have a realization as D-brane world volume gauge theories.
In order to show how these special supersymmetric theories are built, we
first review some useful properties on hyperbolic Lie algebras as well as
their algebraic geometry analog. Then, we consider their realization as
world volume gauge theories of D-branes wrapping cycles in CY3s with
hyperbolic geometries.

\subsection{Roots and Homology in Hyperbolic ADE Geometries}

Roughly speaking, there are two main classes of hyperbolic simply laced Lie
algebras: standard hyperbolic algebras and the so called strictly hyperbolic
ones. For precise definitions on these two varieties of indefinite Lie
algebras see \cite{25},\cite{26},\cite{15}. In this paper we will be
interested in standard hyperbolic algebras, they can be thought of as
particular extension of affine simply laced Kac-Moody invariance. Note that
the hyperbolic symmetries we will be considering here are very special, we
will refer to them as hyperbolic ADE algebras, a matter of keeping in mind
that these symmetries contain the usual finite ADE Lie algebras and their
affine analog as Lie subalgebras.

\textit{Roots and cycles in hyperbolic model}

Since hyperbolic ADE Lie algebras are simplest extensions of affine KM ones,
one can easily make an idea on this larger structure. More precisely, basic
tools of these hyperbolic ADE Lie algebras are quite same as those used in
affine KM symmetries. Of course there are also novelties, but these can be
extracted by using similarity arguments to be specified later on. Let
comment briefly how these hyperbolic ADE algebras can be obtained from the
underlying affine ones; for a rigorous derivation and details see \cite{21}.

\textit{Simple roots}:\qquad Starting from a rank r affine simply laced
Kac-Moody algebra with the usual simple roots $\alpha _{i},$ with $%
i=0,1,...,r$ where $\alpha _{0}=\delta -\sum_{i=1}^{r}d_{i}\alpha _{i}$ with
$\delta $ being the usual imaginary root and the other $\alpha _{i}$s the
simple real ones, we can write down several relations for hyperbolic ADE Lie
algebras just by using intuition and similarity arguments. The rule is as
follows: Once an affine ADE Cartan matrix $K_{ij}^{0}=\alpha _{i}.\alpha
_{j} $ and corresponding Dynkin diagram $\mathcal{D}^{0}$ are fixed, a
hyperbolic ADE generalization of Cartan matrix and Dynkin graph are obtained
by adding an extra simple root $\alpha _{-1}$ to the game with the property,
\begin{equation}
\alpha _{-1}.\alpha _{-1}=2;\qquad \alpha _{-1}\alpha _{0}=-1;\qquad \alpha
_{-1}\alpha _{i}=0.
\end{equation}%
It happens that, because of the Lorentzian nature of the root lattice of
hyperbolic ADE algebras, this extra simple root is realized in a different
manner comparing to what we do in ordinary and affine cases. In addition to
standard simple roots $\alpha _{i}$, with $i=0,1,...,r$, we have,
\begin{equation}
\alpha _{-1}=\gamma -\delta ;\qquad \gamma ^{2}-2\gamma .\delta +\delta
^{2}=2,
\end{equation}%
with $\gamma $ a second basic imaginary root related to the standard
imaginary root $\delta $ of affine KM as well as to the previous real simple
roots $\alpha _{i}$ of finite subalgebra as follows,
\begin{eqnarray}
{\gamma }^{2} &=&{\delta }^{2}=0;\qquad {\gamma }{\delta }=-1,  \notag \\
{\gamma .}\alpha _{0} &=&1;\qquad {\gamma .}\alpha _{i}=0,\ i>0.
\end{eqnarray}%
Therefore our hyperbolic ADE Lie algebras have two basic imaginary roots $%
\gamma $ and $\delta $ which a priori should play a completely symmetric
role. This property will be one of the main arguments that allow us to
extract part of remaining information on the hyperbolic ADE structure that
are not captured by affine KM subsymmetry. \ Note that the symmetric role
between $\gamma $ and $\delta $ is not hazardous; but it reflects invariance
of hyperbolic root system under Weyl transformations. For details we refer
to the above mentioned paper.

\textit{Cycles}:\qquad In the present study we are also interested in type
IIB string interpretation on CY3s of the 4D supersymmetric quiver gauge
theories based on the Dynkin diagrams of hyperbolic algebras. It is then
interesting to find a geometric meaning of the $\gamma $ and $\delta $
imaginary roots in homology of CY3s with hyperbolic ADE geometry. Formal
analogy with affine models shows that $\gamma $ and $\delta $ should
correspond\ to two homological two cycles of CY3s. Extending results on type
II string interpretation of affine models to the hyperbolic ADE case, we see
that geometrically speaking, $\gamma $ and $\delta $ generate a two torus;
or more precisely two heterotic torii $T_{\pm }^{2}$. The latters may be
thought of as holomorphic T$_{z}^{2}$ and anti-holomorphic T$_{\overline{z}%
}^{2}$ two cycles in Kahler threefolds. This observation constitutes a key
point in our analysis of hyperbolic ADE quiver gauge models. It allows us to
construct the complex two dimension homology of the needed CY3s.

In what follows, we build the explicit root system of hyperbolic ADE Lie
algebras without using involved technical details. The corresponding
hyperbolic algebraic geometry is recovered by using the standard dictionary
on this matter relating Lie algebra roots and algebraic geometry
corresponding quantities.

\textit{Hyperbolic root system}

As noted before, the $\gamma $ and $\delta $ imaginary roots allow us to get
many precious informations not only on hyperbolic Lie algebra, but also on
corresponding algebraic geometry analog. We focus our attention on the
derivation of the root system of hyperbolic ADE Lie algebras from which we
directly determine the corresponding CY3s homological cycles. To that
purpose, note that from construction the imaginary roots $\gamma $ and $%
\delta $ play a completely symmetric role. This symmetric property, which
corresponds to a Weyl rotation, turns out to have important implications on
the construction of hyperbolic ADE root system, the derivation H$_{2}$
homology classes of the CY3 with hyperbolic geometry and on the study of
hyperbolic quiver gauge theories. One of these implications follows from the
fact that as roots, $\gamma $ and $\delta $ can be usually expanded in terms
of the simple roots $\left\{ \alpha _{i};-1\leq i\leq r\right\} $ of the
hyperbolic algebra as,
\begin{equation}
\delta =\sum_{i=0}^{r}d_{i}\alpha _{i};\qquad \gamma
=\sum_{i=-1}^{r}d_{i}\alpha _{i},  \label{torr}
\end{equation}%
where $d_{i}$, $0\leq i\leq r$\ are the usual Dynkin weights and where $%
d_{-1}$, which is equal to one, is their analog for the hyperbolic node. In
the present study, we will think about all these $d_{i}$s ( $-1\leq i\leq r$%
) as generic positive integers and treat them on equal footing. While in
affine KM symmetries, $\delta =\sum_{i=0}^{r}d_{i}\alpha _{i}$ is a standard
relation, $\gamma =\sum_{i=-1}^{r}d_{i}\alpha _{i}$ is\ a specific one for
the hyperbolic case. Note that upon setting $d_{0}=0$ in the expansion of $%
\gamma $, we see that $\gamma ^{\ast }=\gamma |_{d_{0}=0}$ can play
perfectly the role of a KM imaginary root in the same manner as does $\delta
$. A way to see this property is to rename the $\alpha _{-1}$simple root as $%
\alpha _{-1}|_{d_{0}=0}=\alpha _{0}^{\ast }$ and notes that the restriction
( projection) $\delta |_{d_{0}=0}$ is just the maximal root $\psi $ of the
usual underlying finite dimensional Lie algebra. Moreover in homology
language the above relations correspond to the two cycles,
\begin{equation}
T_{+}^{2}=\sum_{i=0}^{r}d_{i}S_{i}^{2};\qquad
T_{-}^{2}=\sum_{i=-1}^{r}d_{i}S_{i}^{2},  \label{tor}
\end{equation}%
where naively the $S_{i}^{2}$s are the usual real two spheres of type II
strings on CY3s with deformed ADE singularities. Using results of affine KM
invariance and the symmetric role between $\gamma $ and $\delta $, we can
derive the hyperbolic root ( two cycles) system $\Delta _{hyp}$ (in CY3s).
Following \cite{21}, the full set $\Delta _{hyp}$ of roots of hyperbolic Lie
algebras is as follows,
\begin{equation}
\Delta _{hyp}\cup \{0\}=\{a=n\gamma +m\delta +l\alpha ;n,m\in Z;l=0,\pm
1;\alpha \in \Delta _{finite}\},  \label{al}
\end{equation}%
where $\Delta _{finite}$ stands for the roots system of finite dimensional
ADE Lie sub-algebras. A rigorous way to get the system $\Delta _{hyp}$ is to
solve constraint eqs required by the hyperbolic structure \cite{21}.
Nevertheless, there is also short path to get this result. The idea, which
is done in two steps, is as follows:

(\textbf{1}) Observe that our hyperbolic Lie algebra $g_{hyp}$ has
remarkably two special affine Kac-Moody sub-algebras $g_{aff}^{\gamma }$ and
$g_{aff}^{\delta }$ in one to one correspondence with the two imaginary
roots $\gamma $ and $\delta $ encountered above. So root system $\Delta
_{hyp}$ should contain as subsystems $\Delta _{aff}^{\gamma }$ and $\Delta
_{aff}^{\delta }$ respectively generated by $\gamma $ and $\delta $. This is
clearly seen on above relation from which it is not difficult to see that
the root systems of $g_{aff}^{\gamma }$ and $g_{aff}^{\delta }$ are
respectively given by,
\begin{eqnarray}
\Delta _{aff}^{\gamma }\cup \{0\} &=&\{a=n\gamma +l\alpha ;n\in Z;l=0,\pm
1;\alpha \in \Delta _{finite}\},  \notag \\
\Delta _{aff}^{\delta }\cup \{0\} &=&\{a=m\delta +l\alpha ;m\in Z;l=0,\pm
1;\alpha \in \Delta _{finite}\},  \label{al1}
\end{eqnarray}%
where $\gamma $ and $\delta $ play the role of the basic imaginary root in
each sector.

(\textbf{2}) Think about hyperbolic Lie algebra just as an interpolation
between these two affine sub-symmetries where roots $n\gamma +m\delta
+l\alpha $ are between $n\gamma +l\alpha $ and $m\delta +l\alpha $ recovered
by solving the constraint eq $nm=0$. The special case $m=n=0$ corresponds
just to root system of underlying finite dimensional ADE. This way of doing
can be also used to derive Weyl group $W_{hyp}$, it\ is given by $%
W_{finite}\ltimes T$; i.e the semi-direct product of ordinary Weyl group $%
W_{finite}$ with the group of translations $T$ in the hyperbolic co-root
lattice.

\subsection{Hyperbolic gauge model}

As already noted, hyperbolic quiver gauge theories can be thought of as the
simplest extension of supersymmetric affine ADE ones. All what we know on
affine quiver gauge models can be extended to the hyperbolic case. Let us
consider the extension of those quantities that are relevant for the present
study by using the similarity argument.

To make a parallel between hyperbolic and affine supersymmetric quiver
theories as well as their type II string interpretations, one has to answer
a basic question regarding string coupling $g_{s}$. As we know, this
coupling is strongly linked with the imaginary root $\delta $ ( two torus)
of the deformed elliptic singularity of CY3 on which type II string is
compactified. In hyperbolic quiver gauge models, one has two fundamental
imaginary roots $\gamma $ and $\delta $ ( two heterotic torii) and so one
should expect two kinds of basic moduli. This is not a problem in type IIB
string since we have two candidates for these moduli namely the dilaton
field $\phi $ and the RR axion $\chi $. In addition to type II string
compactification on CY3s with elliptic singularities, we see that we also
have solutions involving CY3 folds with hyperbolic singularities. A way to
motivate this class of backgrounds is to note that the usual type IIB string
coupling constant,%
\begin{equation}
g_{s}=exp\phi ,  \label{gs}
\end{equation}%
can be usually promoted to $g_{\pm }$ by considering non zero values for the
RR\ axion field $\chi $. This may be done by mimicking the construction used
in the derivation of F-theory. The idea consists to shift string coupling $%
g_{s}$ as follows,
\begin{equation}
g_{s}^{-1}=exp(-\phi )\qquad \rightarrow \qquad g_{\pm }^{-1}=exp(-\phi )\pm
\chi ,  \label{g+-}
\end{equation}%
but still keeping reality condition. This leads however to an indefinite
product $g_{+}g_{-}$ since the quantity $g_{+}^{-1}g_{-}^{-1}=g_{s}^{-2}-%
\chi ^{2}$ has a Lorentzian signature. Upon performing a Wick rotation of
the RR axion moduli ($\chi \longrightarrow i\chi $), the inverse string
coupling $g_{+}^{-1}$ becomes $g_{\tau }^{-1}=exp(-\phi )-i\chi $ which
should be compared with the complex moduli $i\tau _{IIB}$ of the holomorphic
two torus used in F-theory compactification on elliptic K3 \cite{26}. The
other coupling $g_{-}^{-1}$ becomes $g_{\overline{\tau }}^{-1}=exp(\phi
)+i\chi $ and should be associated with the complex moduli $-i\bar{\tau}%
_{IIB}$ of the anti-holomorphic torus obtained by taking complex
conjugation. Moreover demanding positivity of the $g_{\pm }$ couplings
constants oblige the range of $\phi $ and $\chi $ moduli to satisfy a
constraint. Taking the dilaton as the basic modulus with an arbitrary real
value, the range of the axion RR field should like,
\begin{equation}
\left\vert \chi \right\vert \leq exp(-\phi ).
\end{equation}%
To get the relation between $g_{\pm }$ and the $g_{i}$ moduli of the
hyperbolic quiver gauge theory, it is enough to recall the relation between
the string coupling $g_{s}$ and the $g_{i}$ gauge couplings in the affine
model.

\begin{equation}
g_{s}^{-1}=\sum_{i=0}^{r}d_{i}g_{i}^{-2},
\end{equation}
where $d_{i}$s are the Dynkin weight encountered above. This relation can be
viewed as just corresponding to the special case of a zero axion field. For
the general case where there is a non zero axion, we have two couplings $%
g_{\pm }$ and so the previous relation generalizes as follows:

\begin{equation}
g_{-}^{-1}=\sum_{i=0}^{r}d_{i}g_{i}^{-2};\qquad
g_{+}^{-1}=\sum_{i=-1}^{r}d_{i}g_{i}^{-2};
\end{equation}%
where $g_{-1}$ is the gauge coupling of the gauge group engineered on the
hyperbolic node of the Dynkin diagram $\mathcal{D}^{-}$. From above
relation, we also see that the parameters $g_{\pm }$ are linked as follows,

\begin{equation}
g_{+}^{-1}=g_{-1}^{-2}+g_{-}^{-1},
\end{equation}%
Substituting $g_{+}^{-1}$ and $g_{-}^{-1}$ by the dilaton and the axion
expressions, we get $g_{-1}^{-2}=2\chi $. This show that the RR axion field
is identified with the volume of the hyperbolic node in the hyperbolic ADE
geometry. Taking $\chi =0$, one recovers the usual affine model with all
desired features.

\subsection{Brane realization}

A way to get the D-brane realization of hyperbolic ADE gauge theories living
in D-brane world volume is as follows. Start from the usual D-brane
representation of the affine ADE quiver gauge model and appropriately deform
the gauge system towards the hyperbolic case. This can be done by help of
the root system of the hyperbolic ADE algebras and their algebraic geometry
analog.

\textit{Affine model}

To begin recall that in affine ADE model, one has in general $N_{0}$
coincident regular D3-branes and $N=\sum_{i=1}^{r}N_{i}$ fractional
D5-branes. Each block of the $r$ sets of $N_{i}$ D5-branes is wrapped over
two cycles of the deformed ADE geometry of the local CY3. The homological
properties of these $r$ real two cycles can be obtained from the properties
of simple roots of ADE Lie algebra. One of these features is that the
irreducible components $\mathbb{P}_{i}^{1}$ of the homological chain $%
\sum_{i}d_{i}\mathbb{P}_{i}^{1}$ are not free but rather glued as follows,
\begin{equation}
C_{\psi }=\sum_{i=1}^{r}d_{i}\mathbb{P}_{i}^{1};\qquad C_{\psi
}^{2}=-2;\qquad \mathbb{P}_{i}^{1}.\mathbb{P}_{j}^{1}=-K_{ij}^{+}.
\end{equation}%
In this relation the two cycle $C_{\psi }$ is the open two-chain
corresponding to the maximal root $\psi =\sum_{i}d_{i}\alpha _{i}$ of finite
ADE Lie algebra. In fact the $C_{\psi }$ chain should be thought of as just
a part of the following closed two-chain
\begin{equation}
C_{\delta }=\sum_{i=0}^{r}d_{i}\mathbb{P}_{i}^{1}=C_{0}+C_{\psi };\qquad
\mathbb{P}_{i}^{1}.\mathbb{P}_{j}^{1}=-K_{ij}^{0};\qquad C_{\delta }^{2}=0,
\label{tor1}
\end{equation}%
which is associated with the imaginary root $\delta $\ of affine ADE
Kac-Moody algebras. In the above eqs, $K_{ij}^{+}$ and $K_{ij}^{0}$\ are as
before, they stand for the usual ordinary and affine ADE Cartan matrices
respectively. From these eqs, it is not difficult to see that $N_{0}$ is the
number of regular D-branes filling space time and living on the loop $%
C_{\delta }$. Therefore the gauge group of the $\mathcal{N}=2$
supersymmetric affine ADE quiver QFT$_{4}$ is,
\begin{equation}
\prod_{i=0}^{r}U\left( \widehat{N}_{i}\right) =U\left( N_{0}\right) \times
\prod_{i=1}^{r}U\left( N_{i}+d_{i}N_{0}\right)
\end{equation}%
where for convenience we have set the $\widehat{N}_{i}$s ranks as \cite{13},
\begin{equation}
\widehat{N}_{0}=N_{0};\qquad \widehat{N}_{i}=N_{0}d_{i}+N_{i};\qquad 1\leq
i\leq r.
\end{equation}%
Knowing the gauge group, one can go ahead and study the quantum dynamics of
these supersymmetric quiver gauge theories. Of particular interest is the
beta functions $\beta _{i}=\beta \left( g_{i}\right) $ for the gauge
couplings $g_{i}$ and the RG flows $\beta _{i}=\beta _{i}\left( \mu \right) $%
. We will turn to this aspect after giving the brane realization for
hyperbolic gauge models.

\textit{Hyperbolic Model}

In hyperbolic quiver gauge theories extending the previous affine ADE ones,
we have also regular coincident D3-branes and fractional D5 ones. But now
with a quiver gauge group given by,
\begin{equation}
\prod_{i=-1}^{r}U\left( \widehat{N}_{i}\right) =U\left( N_{-1}\right) \times
U\left( N_{0}+d_{0}N_{-1}\right) \times \prod_{i=1}^{r}U\left(
N_{i}+d_{i}\left( N_{-1}+N_{0}\right) \right) ,
\end{equation}%
that is one factor more and an extra stuck of $N_{-1}$ D3-branes. The point
is that in case of hyperbolic symmetries, the underlying geometry involves
two heterotic torii $C_{\delta }$ and $C_{\gamma }$. The first one is same
as the two torus $C_{\delta }$ of eq(\ref{tor1}), which we call $C^{-}$ and
the second is the new closed heterotic chain $C_{\gamma }\equiv C^{+}$ built
as,
\begin{equation}
C^{+}=C^{-}+\mathbb{P}_{-1}^{1};\qquad C^{\pm }.C^{\pm }=0;\qquad
C^{-}.C^{+}=1.
\end{equation}%
Mimicking the analysis we have done for affine ADE models, we immediately
get the D-brane representation of hyperbolic ADE quiver gauge theories. The
D-brane system realizing hyperbolic quiver gauge model depends on the sign
of the axion modulus and is as follows:

(a) \textit{Minimal realization (}$\chi <0$\textit{)}:\qquad Besides the
previous $N_{0}$ D3-branes on $C^{-}$ and the wrapped $N$ D5 ones ($%
N=\sum_{i=1}^{r}N_{i}$), we have moreover an extra subset of $N_{-1}$
D3-branes filling the four space time and living on $C^{+}$. These $N_{-1}$
D3-branes capture the hyperbolic extension; and though they look like $N_{0}$
D3-branes from space time view, they carry different internal quantum
numbers as they live on a different internal geometry cycle. Therefore
D-branes of hyperbolic quiver gauge theories have brane charges $\widehat{N}%
_{i}$ given by,
\begin{equation}
\widehat{N}_{-1}=N_{-1};\qquad \widehat{N}_{0}=N_{0}+d_{0}N_{-1};\qquad
\widehat{N}_{i}=\left( N_{0}+N_{-1}\right) d_{i}+N_{i};\qquad 1\leq i\leq r.
\label{repn}
\end{equation}%
These numbers, which can be also read from the algebraic relation (\ref{torr}%
), are direct extension of the affine ADE model considered before. Note that
setting $N_{-1}=0$, one recovers the standard affine model and setting $%
N_{0}=0$ one gets the second supersymmetric quiver CFT$_{4}$ hidden in the
hyperbolic model; see eqs(\ref{al}-\ref{al1}). It would be interesting to
deepen the study which link the two critical behaviours inside hyperbolic
model.

(b) \textit{Non minimal realization (}$\chi >0$\textit{)}:\qquad Along with
the above representation of the hyperbolic gauge system, we may also have a
set of $M_{-1}$ fractional D5-branes wrapping the two cycle
\begin{equation}
C^{-}-C^{+}
\end{equation}%
In this case, eq(\ref{repn}) are replaced as
\begin{equation}
\widehat{N}_{-1}=d_{-1}N_{-1}+M_{-1}  \label{repn1}
\end{equation}%
and the other $\widehat{N}_{i}$\ as before. Note that since wrapping
D5-branes over two cycles involve positive volumes, non zero $M_{-1}$
requires positive axion moduli, $\chi >0$.

\section{RG flows in Hyperbolic model}

To fix the ideas, consider the hyperbolic ADE quiver gauge model described
by the action (\ref{act}) with all real FI couplings $\zeta _{i}$ set to
zero ( $\zeta _{i}=0$). Then focus on linear adjoint matter self-couplings $%
W\left( \Phi _{i}\right) =z_{i}\Phi _{i}$ so that the gauge theory has $%
\mathcal{N}=2$ supersymmetry.

\subsection{Running couplings}

Recall that in the special case where there is no Kahler deformation, the
gauge couplings $g_{i}\sim \sqrt{\zeta _{i}^{2}+z_{i}z_{i}^{\ast }}$ of the $%
\prod_{j=0}^{r}U\left( \widehat{N}_{j}\right) $ symmetry reduce to
\begin{equation}
8\pi ^{2}{g_{i}^{-2}}=\left\vert z_{i}\right\vert ^{2}\geq 0,  \label{gi1}
\end{equation}%
where $\left\vert z_{i}\right\vert =\sqrt{z_{i}z_{i}^{\ast }}$. Since these $%
z_{i}$s can be varied freely in the moduli space of the hyperbolic quiver QFT%
$_{4}$, the corresponding complex deformations induce variations of the ${%
g_{i}^{-2}}$s. These variations have a quantum field theoretic
interpretation in terms of renormalization group flows ${g_{i}=g_{i}}\left(
\mu \right) $. But before going into details, note that for $\zeta _{i}=0$,
the real gauge moduli ${g_{i}^{-2}}$ can be usually split as the product of
holomorphic factor $\QTR{sl}{G}{_{i}^{-1}}=z_{i}\sqrt{\frac{1}{8\pi ^{2}}}$
and its complex conjugate as ${g_{i}^{-2}}=\QTR{sl}{G}{_{i}^{-1}}\left(
\QTR{sl}{G}{_{i}^{-1}}\right) ^{\ast }$. This is an expected feature in
absence of Kahler deformations; it reflects just holomorphy of complex
deformations in $\mathcal{N}=2$ supersymmetric gauge theories. In fact this
splitting is not inherent to $\mathcal{N}=2$ supersymmetry; it is also valid
for $\mathcal{N}=1$ models for which the previous complex gauge couplings $%
\QTR{sl}{G}{_{i}}$ generalize as
\begin{equation}
\QTR{sl}{G}{_{i}^{-1}}=\frac{1}{2\pi \sqrt{2}}W_{i}^{\prime };\qquad
W_{i}^{\prime }=\frac{\partial W}{\partial \Phi _{i}}.
\end{equation}%
Under the change of one or several of the $z_{i}$s in the moduli space of
the hyperbolic quiver QFT$_{4}$, the corresponding ${g_{i}^{-2}}$ gauge
couplings vary too. A physically interesting feature of this change concerns
the case where one of the $z_{i}$s vanishes ( ${z}_{i}=0$). In this
particular situation, the underlying CY3 geometry in type IIB string
compactification becomes singular and the corresponding supersymmetric QFT$%
_{4}$ confine at some gauge scale $\Lambda _{i}$. This property has a
quantum field theoretic description in terms of running coupling constants ${%
g_{i}^{-2}=g_{i}^{-2}}\left( \mu \right) $, which read at ( $\mathcal{N}=2$
exact) one loop as,
\begin{equation}
g_{i}^{-2}\left( \mu \right) =\frac{\beta _{i}}{8\pi ^{2}}\ln \left( \frac{%
\mu }{\Lambda _{i}}\right) ;\qquad -1\leq i\leq r,  \label{uv}
\end{equation}%
where $\beta _{i}=-2\pi \mathrm{i}\beta \left( \tau _{i}\right) $ is the
usual holomorphic beta functions with $2\pi \mathrm{i}\tau _{j}=-8\pi
^{2}g_{j}^{-2}+\mathrm{i}\vartheta _{j}$ and $\Lambda _{i}$ is the gauge
confinement scale we referred to above. Comparing eqs(\ref{gi1}) and (\ref%
{uv}), we get the following relation between the $z_{i}$\ moduli and the
running scale $\frac{\mu }{\Lambda _{i}}$,
\begin{equation}
\left\vert z_{i}\right\vert =\beta _{i}\ln \left( \frac{\mu }{\Lambda _{i}}%
\right) ,  \label{zi}
\end{equation}%
linking hyperbolic ADE singularity in the CY3 of the IIB string
compactification with the quantum field scale $\mu =\Lambda _{i}$. Note that
$\left\vert z_{i}\right\vert $ positivity condition, $\left\vert
z_{i}\right\vert =\beta _{i}\ln \left( \frac{\mu }{\Lambda _{i}}\right) \geq
0$, of the gauge coupling constants can be solved in two ways according to
the value of the renormalization scale $\mu $ and the parameter $\Lambda
_{i} $. We have:

(\textbf{a}) $\beta _{i}>0$\ and $\ln \left( \frac{\mu }{\Lambda _{i}}%
\right) >0$\ which requires $\mu >\Lambda _{i}$, or

(\textbf{b}) $\beta _{i}<0$\ and $\ln \left( \frac{\mu }{\Lambda _{i}}%
\right) <0$ requiring $\mu <\Lambda _{i}$.

The singular limit $\left\vert z_{i}\right\vert =0$ corresponds either to $%
\mu =\Lambda _{i}$ or to $\beta _{i}=0$ independently of the renormalization
group scale's value. At this point, the hyperbolic gauge theory confines and
its interpretation as D-branes world volume is sensitive to the sign of the
beta functions which we propose to discuss now.

\subsection{Holomorphic Beta functions}

Following \cite{15}, the holomorphic beta function factors $\beta _{i}=\beta
\left( g_{i}\right) $ of $\mathcal{N}=2$ supersymmetric $\Pi
_{i=-1}^{r}U\left( \widehat{N}_{i}\right) $\ quiver gauge theories depend
linearly on the $\widehat{N}_{i}$ ranks. They are proportional to the Cartan
matrix $K_{hyp}^{-}$ of the underlying hyperbolic quiver symmetry; $\beta
_{i}\sim K_{ij}^{-}\widehat{N}_{j}$. For an explicit treatment, we will
first consider the particular example of hyperbolic \textbf{A}$_{r}$ QFT$%
_{4} $. The results for hyperbolic DE cases are given in appendix B.

\textbf{Hyperbolic A}$_{r}$\textbf{\ model}:\qquad In the example of $%
\mathcal{N}=2$ supersymmetric hyperbolic A$_{r}$ model, the beta functions
for the quiver gauge group can be easily determined. They read as;
\begin{eqnarray}
\beta _{-1} &=&2\widehat{N}_{-1}-\widehat{N}_{0};\qquad \beta _{0}=-\widehat{%
N}_{-1}+2\widehat{N}_{0}-\widehat{N}_{1}-\widehat{N}_{r},  \notag \\
\beta _{i} &=&2\widehat{N}_{i}-\widehat{N}_{i-1}-\widehat{N}_{i+1};\qquad
1\leq i\leq r-1,\qquad \beta _{r}=2\widehat{N}_{r}-\widehat{N}_{r-1}-%
\widehat{N}_{0}.  \label{beta0}
\end{eqnarray}
where $\widehat{N}_{i}$ are as in eq(\ref{repn}) and (\ref{repn1}) according
to the sign of the axion modulus.

$(i)$ \textit{Minimal representation}:\qquad Up on using eqs(\ref{repn})
defining the minimal realization of the hyperbolic gauge theory, the
previous relations simplify to,
\begin{eqnarray}
\beta _{-1} &=&N_{-1}-N_{0};\qquad \beta _{0}=-N_{-1}-N_{1}-N_{r};\qquad
\beta _{1}=2N_{1}-N_{2},  \notag \\
\beta _{r} &=&2N_{r}-N_{r-1};\qquad \beta _{i}=2N_{i}-N_{i-1}-N_{i+1};\qquad
2\leq i\leq r-1,  \label{beta1}
\end{eqnarray}%
where one recognizes the usual beta functions for finite dimensional A$_{r}$
quiver gauge model. One also has the usual $\beta _{0}$ which is independent
of the $N_{0}$ number of D3-brane living on the C$^{-}$ heterotic cycle but
does depend on the number $N_{-1}$ of D3-branes living on the C$^{+}$
heterotic cycle. This is a key point which turn out to have a drastic
consequence on conformal invariance. Finally $\beta _{-1}$ is proportional
to the difference between the numbers of regular branes living on the two
heterotic torii. In addition to these quantities, we have the following
extra relations,
\begin{equation}
\beta _{\psi }=\left( \beta _{1}+...+\beta _{r}\right) ;\qquad \beta
_{\delta }=\beta _{0}+\beta _{\psi };\qquad \beta _{\gamma }=\beta
_{-1}+\beta _{\delta },
\end{equation}%
where $\beta _{\delta }$ and $\beta _{\gamma }$ are the holomorphic beta
functions of the diagonally embedded groups $U_{D}\left( N_{-1}\right) $ and
$U_{D}\left( N_{0}\right) $ respectively. These groups are in one to one
with the closed two chains of two cycles $C^{\pm }$ ; they are also in one
to one with the two basic imaginary roots $\delta $ and $\gamma $. The
factor $\beta _{\psi }$ is the beta function associated with the maximal
root $\psi =\sum_{i=1}^{r}\alpha _{i}$ of affine A$_{r}$\ algebra. Before
proceeding further, note that in this hyperbolic A$_{r}$ model, we have the
following results:

(\textbf{a}) The beta function $\beta _{0}$ of hyperbolic A$_{r}$ theory is
negative. This shows that the confining scale $\Lambda _{0}$ associated with
the running coupling $g_{0}\left( \mu \right) $ should tend to infinity in
perfect agreement with the string interpretation of quiver gauge theories.

(\textbf{b}) From the value of $\beta _{\psi }=N_{1}+N_{r}$ it follows that
\begin{equation}
\beta _{\delta }=-N_{-1}
\end{equation}%
is negative definite contrary to the affine case where it is null. This
clearly shows that\ even in the absence of wrapped D5-branes, hyperbolic
quiver gauge theory is not conformal invariant in opposition to affine model.

(\textbf{c}) Because of symmetric role played by $\delta $ and $\gamma $, we
should expect here also that
\begin{equation}
\beta _{\gamma }=-N_{0}.
\end{equation}%
This is exactly what we find by substituting the value $\beta
_{-1}=N_{-1}-N_{0}$ into $\beta _{\gamma }=\beta _{-1}+\beta _{\delta }$.
Note that for the special case $N_{-1}=N_{0}$, the factor $\beta _{-1}$
vanishes but $\beta _{0}=-N_{-1}-N_{1}-N_{r}$ vanishes only if $%
-N_{-1}-N_{1}-N_{r}=0$ which requires $N_{-1}=N_{1}=N_{r}=0$. The vanishing
of the other $\beta _{i}$\ factors require furthermore $N_{i}=0$.

So hyperbolic theories as realized above are not conformal invariant.

$(ii)$ \textit{Non minimal representation}:\qquad As we noted before, this
concerns models with positive values of the axion field. In this case, one
may also have $M_{-1}$ D5-branes wrapping the hyperbolic irreducible cycle
and the beta functions given by eqs(\ref{beta1}) are now modified to,
\begin{eqnarray}
\beta _{-1} &=&2M_{-1}+N_{-1}-N_{0};\qquad \beta
_{0}=-M_{-1}-N_{-1}-N_{1}-N_{r};;\qquad \beta _{1}=2N_{1}-N_{2},  \notag \\
\beta _{r} &=&2N_{r}-N_{r-1};\qquad \beta _{i}=2N_{i}-N_{i-1}-N_{i+1};\qquad
2\leq i\leq r-1,
\end{eqnarray}%
This extension can no longer have $\beta _{0}=0$ as far as D-brane world
volume gauge theories are concerned. A part the shift of $\beta _{-1}$, the
introduction of extra $M_{-1}$ D5-branes does bring nothing new.
Consequently, we fix our attention, in what follows, on the study of minimal
representation ($M_{-1}=0$) of hyperbolic theories. To do so, let us begin
by making some comments on hyperbolic brane system having no D5-brane at
all, $M_{-1}=0$ and $N_{i}=0$, $i=1,...,r$.

\textbf{Comments on hyperbolic D3-Brane system}:\qquad This is a very
special representation since there is no fractional D5-branes in the
hyperbolic quiver gauge model. Therefore the set of beta functions $\beta
_{i}$ with $i=1,..r$ are identically zero ( $\beta _{i}=0$) and those
associated with the hyperbolic node and affine one are as follows,
\begin{eqnarray}
\beta _{-1} &=&N_{-1}-N_{0};\qquad \beta _{0}=-N_{-1};\qquad \beta
_{i}=0;\qquad i=1,...,r,  \notag \\
\beta _{\delta } &=&-N_{-1};\qquad \beta _{\gamma }=-N_{0}.
\end{eqnarray}%
Since there is no non trivial solution for the vanishing of these beta
functions, this particular hyperbolic gauge theory can never be made
conformal contrary to the underlying affine model corresponding to taking $%
N_{-1}=0$. In addition to this property, the D3 system realizing the
hyperbolic model exhibits other interesting features which we comment here
below.

\textbf{(a) Signs of} $\beta _{i}$:\qquad First note that as far as $%
N_{-1}\neq 0$, the gauge factor $U\left( \widehat{N}_{0}\right) $ with beta
function $\beta _{0}$ is not asymptotically free and so the corresponding
gauge degrees of freedom are confined. For the case $0<N_{-1}<N_{0}$, the
situation is worst as the gauge degrees of freedom of the gauge group factor
$U\left( \widehat{N}_{-1}\right) $ associated with the hyperbolic node are
confined as well. However for $N_{-1}>N_{0}$, the beta function $\beta
_{-1}>0$ and so the gauge group $U\left( \widehat{N}_{-1}\right) $ is
asymptotically free. Since the beta functions for $U\left( \widehat{N}%
_{i}\right) $ with $1\leq i\leq r$ are null ($\beta _{i}=0$ ), this
situation is very special in the sense that it formally resembles to the $%
\mathcal{N}=1$ supersymmetric Klebanov Strassler affine A$_{1}$ model.
There, beta functions with opposite signs is the necessary condition for the
existence of RG cascades. In the pure D3 system realizing hyperbolic
theories, we have here also $\beta _{-1}$ and $\beta _{0}$\ with opposite
signs as shown below,
\begin{equation}
\beta _{-1}>0>\beta _{0},\qquad \beta _{i}=0;\qquad i=1,...,r.
\end{equation}%
This behavior can be viewed as interesting indication for existence of RG
cascades in hyperbolic gauge theories. To avoid confusion in what follows,
we will refer to these beta functions as $\beta _{i}^{\left( 0\right) }$, $%
-1\leq i\leq r$ where the upper index refers to the fact that we are dealing
with the original theory. More generally, we denote the beta functions of
the \textit{n-th step }of\textit{\ }the RG cascades as\textit{\ }$\beta
_{i}^{\left( n\right) }$.\textit{\ }

\textbf{(b)} \textbf{Dual models}:\qquad Using Seiberg like duality
transformation idea and following \cite{13}, we can dualize the above
hyperbolic theory as usual by using Weyl transformations of hyperbolic A$%
_{r} $ Lie algebra. The previous beta functions get now replaced by their
Weyl transformed ones,
\begin{eqnarray}
\beta _{-1}^{\left( 1\right) } &=&-N_{0};\qquad \beta _{0}^{\left( 1\right)
}=N_{-1},  \notag \\
\beta _{1}^{\left( 1\right) } &=&\beta _{r}^{\left( 1\right)
}=-N_{-1};\qquad \beta _{i}^{\left( 1\right) }=0,\qquad i=2,...,r-1.
\end{eqnarray}%
As we see, these beta functions have changed their signs ( $\beta
_{1}^{\left( 1\right) },\beta _{r}^{\left( 1\right) }<0<\beta _{0}^{\left(
1\right) }.$). The initial gauge group $G_{hyp}^{^{\left( 0\right)
}}=U\left( N_{-1}\right) \times U\left( N_{-1}+N_{0}\right) ^{r+1}$ of the
original regular quiver gauge model is now mapped to,%
\begin{equation}
G_{hyp}^{^{\left( 1\right) }}=U\left( N_{-1}\right) \times U\left(
2N_{-1}+N_{0}\right) \times U\left( N_{-1}+N_{0}\right) ^{r},
\end{equation}%
which can determine us the gauge and matter content of the dual gauge
theory. By performing the next dualization on the node $\beta _{1}^{\left(
1\right) }$, we get the second\ dual generation. Its gauge group is $%
G_{hyp}^{^{\left( 2\right) }}=U\left( N_{-1}\right) \times U\left(
2N_{-1}+N_{0}\right) ^{2}\times U\left( N_{-1}+N_{0}\right) ^{r-1}$ and the
corresponding beta functions are
\begin{eqnarray}
\beta _{-1}^{\left( 2\right) } &=&-N_{0};\qquad \beta _{0}^{\left( 2\right)
}=0;\qquad \beta _{1}^{^{\left( 2\right) }}=N_{-1},  \notag \\
\beta _{2}^{\left( 2\right) } &=&\beta _{r}^{\left( 2\right)
}=-N_{-1};\qquad \beta _{i}^{\left( 2\right) }=0,\qquad i=3,...,r-1.
\end{eqnarray}%
Note that we get the same result if one chooses to dualise the node $\beta
_{r}^{\left( 1\right) }$\ instead of $\beta _{1}^{\left( 1\right) }$.

In general, we have for the \textit{n-th} dualization with $n=0,...,r-1$,
\begin{equation}
G_{hyp}^{^{\left( n\right) }}=U\left( N_{-1}\right) \times U\left(
2N_{-1}+N_{0}\right) ^{n}\times U\left( N_{-1}+N_{0}\right) ^{r+1-n}.
\end{equation}%
The beta functions are
\begin{eqnarray}
\beta _{-1}^{^{\left( n\right) }} &=&-N_{0};\qquad \beta _{0}^{^{\left(
n\right) }}=0=\beta _{i}^{\left( n\right) }=0;\qquad i\neq n-1,n,r  \notag \\
\beta _{n-1}^{\left( n\right) } &=&N_{-1};\qquad \beta _{n}^{\left( n\right)
}=\beta _{r}^{\left( n\right) }=-N_{-1}
\end{eqnarray}%
This construction can be also worked out for $n\geq r$.

\subsection{RG Cascades}

Here we consider minimal hyperbolic quiver gauge model and study RG flows
towards infrared. We first illustrate the method on hyperbolic A$_{2}$
example; the results for generic hyperbolic ADE case are reported in
appendix. For other discussions regarding RG cascades in supersymmetric
quiver gauge theories see \cite{18} for affine $\widehat{A}_{1}$ gauge
theory and \cite{27} for the hyperbolic model based on the non simply laced
Cartan matrix,
\begin{equation}
K_{Fiol}=\left(
\begin{array}{cc}
2 & -k \\
-k & 2%
\end{array}%
\right) .
\end{equation}%
In this particular hyperbolic model, $k$ is a positive integer greater than
two.

\textbf{Hyperbolic A}$_{2}$\textbf{\ model:\qquad }First, recall that
hyperbolic A$_{2}$ quiver gauge model as constructed in the present paper
can be viewed as one of possible generalizations of affine A$_{1}$ quiver
gauge theory known to exhibit impressive features including RG cascades.
This behaviour is then a common property shared by all affine supersymmetric
quiver theories, but also shared by their hyperbolic extensions. For affine
case, RG cascades were considered in different occasions \cite{28}, but a
remarkable Lie algebraic proof was given in \cite{13}. There, it was shown
that RG cascades has much to do with translation sub-symmetries of affine
Weyl group $W_{aff}=W_{finite}\ltimes T$. Here, we will show that this
feature is also true for hyperbolic models.

It is interesting to recall the RG cascade's procedure in affine A$_{2}$
quiver gauge model before giving its proof in hyperbolic case.
\begin{figure}[tbh]
\begin{center}
\epsfxsize=10cm \epsffile{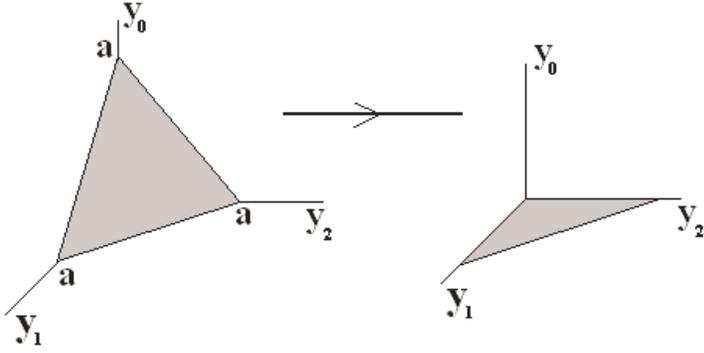}
\end{center}
\caption{\textit{Figure 1a represents the Coxeter box of affine A}$_{2}$%
\textit{\ in the 3-space }$\left( y_{0},y_{1},y_{2}\right) $\textit{. This
is a triangle in 3-space with edges normal to the simple roots. Figure 1b
gives its projection on the }$\left( y_{1},y_{2}\right) $\textit{\ plane
which is also a triangle (a planar triangle).}}
\end{figure}
The general group of quiver gauge symmetry in the UV regime involves
obviously three factors as $U\left( N_{0}\right) \times U\left(
N_{0}+N_{1}\right) \times U\left( N_{0}+N_{2}\right) $ with running gauge
coupling constants $g_{0}$, $g_{1}$ and $g_{2}$ respectively. The factor $%
U\left( N_{0}\right) $\ is engineered on the affine node of the Dynkin
diagram of affine A$_{2}$ and $U\left( N_{0}+N_{1}\right) \times U\left(
N_{0}+N_{2}\right) $ on the two other ordinary nodes. Along with these
factors, there is also a diagonal $U_{D}\left( N_{0}\right) $ sub-symmetry
that turns out to play a central role in the study of RG flows. It deals
with the fundamental imaginary root $\delta $ of affine A$_{2}$ and has a
zero vanishing beta function $\beta _{\delta }$. Its coupling constant is
given by the type IIB string coupling constant $g_{s}$. In \cite{13}, RG
cascades towards infrared has been shown to be accompanied by a decreasing
of the charge of D3-branes ( $\Delta N_{0}<0$). This property has an
interpretation not only in terms of Weyl reflections (Seiberg like duality)
but also as translations in the root lattice generating the affine RG
cascades. The decreasing of D3 charge is easily seen on the Cachazo \textit{%
et al} picture using a formal analogy between flows in supersymmetric quiver
gauge theories and dynamics of a classical particle with coordinates $%
X=\sum_{j=0}^{r}y_{j}\lambda _{j}$, $\alpha _{i}.\lambda _{j}=\delta _{ij}$,
here $\lambda _{j}$s are the fundamental weights of the affine Lie algebra
(affine A$_{2}$). Up to translations, the dynamics of the $X$ particle is
restricted inside the affine A$_{2}$ Coxeter box defined by,
\begin{equation}
g_{i}^{-2}\left( X\right) =\alpha _{i}\cdot X=y_{i}\geq 0,
\end{equation}%
where $\alpha _{i}$s are the usual simple roots of affine A$_{2}$. Note that
this 3-dimensional particle is attached to the walls of the Coxeter box by $%
N_{0}+N_{1}+N_{2}$ strings. The particle is attached on each \textit{i-th}
Coxeter wall by $N_{i}$ strings of equal tension and their length is
obviously given by $g_{i}^{-2}\left( X\right) =y_{i}$. Note also that for
the case where the X particle is on the \textit{i-th} wall, we have $%
g_{i}^{-2}\left( X\right) =0$. So divisors $y_{i}=0$ are in one to one
correspondence with the simple roots $\alpha _{i}$ of the affine $A_{2}$
algebra. In this case, the $X$ vector is normal to $\alpha _{i}$. Figure 1a
gives the Coxeter box for affine $A_{2}$ model in the plane $%
\left(y_{1},y_{2}\right) $.

\begin{figure}[tbh]
\begin{center}
\epsfxsize=6cm \epsffile{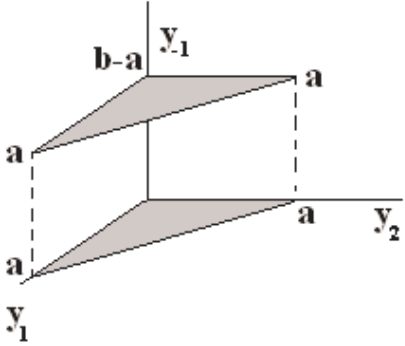}
\end{center}
\caption{\textit{It represents hyperbolic A}$_{2}$\textit{\ Coxeter box}. We
have four variables related as $g_{-}^{-1}=y_{0}+y_{1}+y_{2}$\textit{\ and }$%
g_{+}^{-1}=y_{-1}+g_{-}^{-1}$\textit{, (}$a=g_{-}^{-1},$ $b=g_{+}^{-1}$%
\textit{). Note that }$b-a$ is proportional to axion $\protect\chi $. In the
limit $\protect\chi \rightarrow 0,$ this figure coincides with figure 1b.%
\textit{.}}
\end{figure}

Now using Cachazo et \textit{al} mechanical analogy, RG flows in affine
quiver gauge model have a nice representation as,
\begin{equation}
\frac{dX}{dt}=F,
\end{equation}%
where the force $F$ acting on the particle is given by $F=\sum_{j=0}^{2}%
\alpha _{j}N_{j}$. This equation of evolution is immediately obtained by
using the explicit expression of the running gauge couplings ${g_{i}^{-2}}%
\left( \mu \right) =\frac{\beta _{i}}{8\pi ^{2}}\ln \frac{\mu }{\Lambda _{i}}
$ and setting $8\pi ^{2}y_{i}\left( t\right) =\beta _{i}\left(
t_{i}-t\right) $ with $t=-\ln \mu $ ( $dt=-d\mu /\mu >0$ for flows towards
IR). In this picture Seiberg like duality is nicely described in terms of
Weyl reflections on walls of the Coxeter box. They map $y_{i}$ variables to
\begin{equation}
y_{i}^{\prime }=y_{i}-K_{ij}y_{j},
\end{equation}%
where $K_{ij}$ is the affine $A_{2}$\ Cartan matrix. Geometrically, this is
interpreted as flop transition where a dual $S^{2}$ is blown up. An
analogous description is valid for RG cascades except now they are
associated with Weyl translations. However, in this case, the particle is
diffused outside the fundamental domain of the Coxeter box. Nevertheless
Weyl translations act as shift operators and allow to bring the particle
back inside the fundamental domain. In hyperbolic A$_{2}$ model, the method
is quite similar although the theory is non conformal. The gauge group is $%
U\left( N_{-1}\right) \times U\left( M_{0}\right) \times U\left(
M_{0}+N_{1}\right) \times U\left( M_{0}+N_{2}\right) $, where we have set $%
M_{0}=N_{-1}+N_{0}$, and the corresponding gauge couplings are $g_{-1\text{,}%
}$ $g_{0}$, $g_{1}$ and $g_{2}$ respectively. This hyperbolic model has two
diagonal sub-symmetries $U_{D}\left( N_{-1}\right) $ and $U_{D}\left(
M_{0}\right) $ in one to one with the two fundamental imaginary roots $%
\delta $ and $\gamma $ and the $g_{-}$ and $g_{+}$ couplings eq(\ref{g+-}).
In the UV regime of $\mathcal{N}=2$ hyperbolic A$_{2}$, viewed as a
deformation of affine A$_{2}$ UV theory, one is interested in positive
definite beta functions ensured by requiring,
\begin{equation}
\beta _{1}=2N_{1}-N_{2}>0;\qquad \beta _{2}=2N_{2}-N_{1}>0.
\end{equation}%
These relations fix the range of allowed UV values of part of the gauge
group ranks namely $U\left( M_{0}+N_{1}\right) \times U\left(
M_{0}+N_{2}\right) $. For the other ranks $N_{0}$ and $N_{-1}$, the
situation is a little bit subtle, the $U_{D}\left( N_{-1}\right) $ and $%
U_{D}\left( M_{0}\right) $ gauge factors are no longer conformal. The
corresponding beta functions $\beta _{\gamma }$ and $\beta _{\delta }$ are
negative because $\beta _{\gamma }=-N_{0}$ and $\beta _{\delta }=-N_{-1}$.
But this is not a problem since what interest us in the study of RG flows is
not the diagonal group but rather the gauge factors $U\left( N_{0}\right) $
and $U\left( N_{-1}\right) $ associated with the affine and hyperbolic nodes
respectively. Like in affine case and because of stringy origin, their beta
functions $\beta _{0}$ ( $\beta _{0}=-\left( N_{-1}+N_{1}+N_{2}\right) $)
and $\beta _{-1}$\ ( $\beta _{-1}=N_{-1}-N_{0}$) are not required to be
positive. Therefore, we have to distinguish between different cases
according to signs of these beta functions. Moreover, since $\beta _{0}$ is
negative exactly like in affine case, it is then natural to adopt the same
condition one uses in the affine model namely $N_{0}$ large enough with
respect to the ranks $N_{1}$ and $N_{2}$; i.e,
\begin{equation}
2N_{0}>>N_{1}+N_{2}.  \label{afn}
\end{equation}%
For the rank $N_{-1}$, it is fixed by the sign of $\beta _{-1}$. For $\beta
_{-1}$ positive ( negative) definite, we should have $N_{-1}>N_{0}$ ($%
N_{-1}<N_{0}$) and for $\beta _{-1}=0$ we have $N_{-1}=N_{0}$. So in the UV
regime of hyperbolic theory with positive $\beta _{-1}$, we should have the
following conditions,

\begin{equation}
N_{-1}>N_{0}>N_{1};\qquad N_{-1}>N_{0}>N_{2}.
\end{equation}%
In this case, we dispose at least of two beta functions with opposite signs.

\textbf{Cascades in hyperbolic theory}:\qquad To study RG cascades in
hyperbolic A$_{2}$ quiver gauge model, we have to compute the variations of $%
N_{0}$ and $N_{-1}$ during cascading towards infrared and prove that the
variation of the total charge of D3-branes $M_{0}=N_{0}+N_{-1}$ is negative.
This condition, which extends what we know about cascades in affine quiver
gauge theories, follows also from the fact that in hyperbolic gauge models
the role of $\Delta N_{0}$ of affine models is played by $\Delta M_{0}$ as
can be seen on the structure of the quiver gauge group. These two quantities
should be related with the structure of the Weyl group in hyperbolic theory.
Remember that like in affine case, the Weyl group W$_{hyp}$ of hyperbolic
algebras is also a semi-direct product between reflections W$_{finite}$ and
translations T. The main difference between W$_{hyp}$ and W$_{aff}$\ groups
is that in the former the set of translations is larger.\ This property
follows from the fact that in hyperbolic symmetries one has two basic
imaginary roots $\delta $\ and $\gamma $ and so two fundamental translations
T$_{\alpha }^{\delta }$ and T$_{\alpha }^{\gamma }$. These translations are
indeed the generators of RG cascades in hyperbolic gauge models.

To get the variations of $N_{0}$ and $N_{-1}$, we borrow results from
geometric unification of dualities in ordinary and affine supersymmetric
quiver gauge theories and prove the following statements.
\begin{equation}
\Delta M_{0}<0;\qquad \text{\ but \ }\Delta N_{0}\text{ \ and \ }\Delta
N_{-1}\text{ \ have indefinite sign}
\end{equation}%
To do so we start by recalling the way we get the negative sign of $\Delta
N_{0}$ in $\mathcal{N}=1$ affine quiver gauge models. There, one uses a
geometric approach which expresses $\Delta N_{0}$ as the product of the
fluxes of $H_{NS}$ and $H_{RR}$ field strengths of F1 and D1 strings \cite%
{13,29},
\begin{equation}
\Delta N_{0}=8\pi ^{2}\int_{B_{\delta }}H_{NS}\wedge \int_{A_{\delta
}}H_{RR},  \label{deln}
\end{equation}%
where the NS-NS and RR fields strengths can be usually decomposed as $%
H=\Sigma _{j}\alpha _{j}H^{\left( j\right) }$ and where A$_{\delta }$ and B$%
_{\delta }$ are three cycles to be described in a moment. This method of
doing has an interpretation in terms of geometric transition in the
underlying Calabi-Yau threefold Y$_{0}$ on which type IIB string is
compactified. The point is that while Weyl reflections on walls of the
Coxeter box are associated with Seiberg like duality, Weyl translations
require crossing Coxeter box walls. This may be naively interpreted as
associated with a geometric transition as suggested by geometric unification
of dualities. This means that after transition, the initial variety Y$_{0}$
with an elliptic A$_{2}$ geometry get replaced by a new one denoted Y$%
_{0}^{\prime }$ with dual pairs of three cycles. In field theory language,
this geometric transition corresponds to switching on the non linear adjoint
matter self-interactions $W\left( \Phi _{i}\right) =\sum_{l=1}^{k+1}\xi
_{l}\Phi ^{l}$, so $\mathcal{N}=2$ supersymmetry is partially broken and the
affine quiver gauge theory has $\mathcal{N}=1$\ supersymmetry. Because of
non linearity of $W\left( \Phi _{i}\right) $, the Calabi-Yau threefold Y$%
_{0} $ undergo a geometric transition where the two cycles S$_{i}^{2}$ of
the deformed elliptic singularity get replaced by three cycles involving S$%
_{i}^{3}$s. Moreover D5-branes get replaced by their fluxes. The resulting
generalized conical geometry Y$_{0}^{\prime }$ contains then a particular
pair of dual three cycles A$_{\delta }$ and B$_{\delta }$, the sub-index $%
\delta $ recalls that they have been generated out of the closed two cycle C$%
_{\delta }$ of Y$_{0}$. By integration of the relations,
\begin{equation}
\int_{B_{\delta }}H_{NS}^{\left( i\right) }=\Delta \left( g_{i}^{-2}\right)
=R_{i};\qquad \int_{A_{\delta }}H_{R}^{\left( i\right) }=N_{i},
\end{equation}%
and using the fact that $\Delta \left( g_{i}^{-2}\right) =-\beta _{i}\ln
\frac{\mu _{1}}{\mu _{2}}$\ ($\mu _{1}>\mu _{2}$)\ which in mechanical
analogy can be also put in the form $R_{i}=-aN_{i}$\ \cite{13}, we get $%
\Delta N_{0}=-aN^{2}$ which is negative definite as far as there are
D5-brane charges. Negativity of $\Delta N_{0}$ follows of course from the
fact that eigenvalues of ordinary A$_{2}$ Cartan matrix are positive
definite. In the $\mathcal{N}=1$ hyperbolic quiver gauge model, the
situation is quite similar to the above affine discussion. By switching on
the non linear potential $W$, the initial Calabi-Yau threefold Y$_{-}$ with
hyperbolic geometry ( extending previous Y$_{0}$ having an elliptic
geometry) undergoes a geometric transition leading to Y$_{-}^{\prime }$. The
two cycles S$_{i}^{2}$ of initial Y$_{-}$ get replaced by three cycles S$%
_{i}^{3}$. The resulting generalized conical geometry Y$_{-}^{\prime }$
contain then two pairs of dual closed three cycles $\left( A_{\delta
},B_{\delta }\right) $ and $\left( A_{\gamma },B_{\gamma }\right) $.
Extending the previous affine analysis, we can write down immediately the
variations $\Delta M_{0}$ and $\Delta N_{-1}$ during cascading in hyperbolic
theory. For the value of $\Delta M_{0}$, we have a similar relation as in eq(%
\ref{deln}),
\begin{equation}
\Delta M_{0}=8\pi ^{2}\int_{B_{\delta }}H_{NS}\wedge \int_{A_{\delta
}}H_{RR},  \label{m0}
\end{equation}%
which, by help of previous results, reads as $\Delta
M_{0}=-a\sum_{i,j=0}^{r}K_{ij}N_{i}N_{j}$, $a>0$. Because of the identity $%
\sum_{i,j=0}^{r}K_{ij}d_{i}d_{j}=0$, this is a negative quantity as far as $%
N_{i}$'s are non zero. This proves our first statement. For the second
statement, it is enough to note that the variation $\Delta N_{-1}$ is linked
with the imaginary root $\gamma $. So we have,
\begin{equation}
\Delta N_{-1}=8\pi ^{2}\int_{B_{\gamma }}H_{NS}\wedge \int_{A_{\gamma
}}H_{RR}.
\end{equation}%
The evaluation of this relation is similar to the computation of $\Delta
M_{0}$ (\ref{m0}), except now we have more general sums as shown on the
following explicit expression $\Delta
N_{-1}=-aN^{2}=-a\sum_{i,j=-1}^{r}K_{ij}N_{i}N_{j}$. Since in hyperbolic Lie
algebras, $N^{2}$ has an indefinite sign, we find that the sign of the
variation of $N_{-1}$ during cascades towards infrared is indeed indefinite.
Finally by help of the relation $M_{0}=N_{0}+N_{-1}$, we can compute the
variation of $N_{0}$, we get $\Delta N_{0}=\Delta M_{0}-\Delta N_{-1}$ which
has also an indefinite sign.

\section{Conclusion and Comments}

In this paper, we have studied the classification of four dimensional
supersymmetric quiver gauge theories which can be embedded in type II
strings on Calabi-Yau threefolds. We have shown that there are three basic
classes in one to one with Vinberg classification theorem of Lie algebras.
We have given their D-brane realizations and analyzed the corresponding RG
flows towards infrared. Among our results, we quote the following:

(1) Along with supersymmetric ordinary and affine ADE quiver gauge models,
there is an \textit{indefinite class} associated with the \textit{indefinite
sector} of Lie algebras sharing most features of usual ordinary and affine
supersymmetric quiver gauge models. Except few tentatives for extensions
such the one on duality walls \cite{27,28}, see also \cite{30}, this
indefinite sector has not been explored before. This is mainly due to the
lack of complete Lie algebraic results in this matter. However despite this
difficulty, we have shown that one can make a quite precise idea about the
structure of indefinite quiver gauge theories by looking at their
interesting hyperbolic subset. The latter which may be thought of as
simplest generalizations of affine models, can be used as a laboratory to
establish characteristic results of indefinite gauge theories.

(2) As far as hyperbolic models are concerned, we have shown that these
supersymmetric gauge theories have very special features. They can be viewed
as low energy limit of type IIB string on CY3s with hyperbolic ADE geometry
and are, surprisingly, strongly linked with the RR axion $\chi $. They have
D-brane realizations depending on the sign of $\chi $ modulus interpreted as
the volume of the cycle $S_{\chi }^{2}$ associated with the hyperbolic node.
According to the signs of $\chi $, we have distinguished three kinds of
models:\newline
(i) Minimal hyperbolic model\textit{\ }associated with $\chi <0$. The word
minimal means that there is no D5-brane wrapping the hyperbolic two cycle.%
\newline
(ii) Non minimal hyperbolic model with $\chi >0$ and partially wrapped
D5-branes on $S_{\chi }^{2}$.\newline
(iii) Finally models with $\chi =0$ coinciding exactly with the usual affine
quiver gauge theory.

(3) We have studied RG cascades in hyperbolic quiver gauge theories. We have
shown that they are also linked with Weyl translations generating a part of
automorphisms of the root system of hyperbolic Lie algebras as in case of
affine gauge models. We have found that during cascading towards infrared
the variations of the D3-brane charges $\Delta N_{0}$ and $\Delta N_{-1}$
have indefinite signs but their sum $\Delta N_{0}+\Delta N_{-1}$ giving the
total D3-brane charge is negative. This result has been derived by help of
the mechanical analogy of \cite{13}, but might be done as well by using the
Klebanov-Strassler and Fiol procedures.

On the other hand and as far as supersymmertic field theory is concerned, we
believe that the field theoretical results we have got for hyperbolic model
such as computation of beta functions and cascades are valid as well for
supersymmetric indefinite quiver gauge theories. Moreover and despite that
these are non conformal, we have shown that hyperbolic, and more generally
indefinite, quiver gauge theories share features we have in ordinary and
affine cases. For instance, starting from a $\mathcal{N}=2$ hyperbolic
(indefinite) gauge model, we can build $\mathcal{N}=1$ ones from geometric
transition scenario. In field theoretical approach, it is enough to switch
on the non linear adjoint matter self-interactions $W\left( \Phi _{i}\right)
=\sum_{l=1}^{k+1}\xi _{l}\Phi ^{l}$ with mass scales $W_{i}^{\prime \prime
}=\Delta $ and proceed following \cite{11,13}. One can also study RG flows
\`{a} la Klebanov-Strassler by computing the $\mathcal{N}=1$ physical beta
functions $\beta _{i}^{phys}$. For running energies $\mu $ larger than
scales $\Delta $, the superpotentials may be neglected and the $\mathcal{N}%
=1 $ one loop holomorphic beta functions remain quite same as in $\mathcal{N}%
=2$ case,
\begin{equation}
\beta _{i}^{N=1}\sim \beta _{i}^{N=2};\qquad \Lambda _{i}^{\mathcal{N}%
=1}\sim \Lambda _{i}\qquad \mu >>\Delta .
\end{equation}%
Below $\Delta $, adjoint matter acquires mass and the $\mathcal{N}=1$
holomorphic beta functions $\beta _{i}^{low}$ as well as the corresponding
gauge scales $\Lambda _{i}^{low}$ get modified as,
\begin{equation}
\beta _{i}^{low}=N_{i}+\beta _{i};\qquad \ln \left( \Lambda
_{i}^{low}\right) =\frac{\beta _{i}}{\beta _{i}^{low}}\ln \left( \Lambda
_{i}\right) +\frac{N_{i}}{\beta _{i}^{low}}\ln \left( \Delta \right) .
\end{equation}%
To get the RG flows in $\mathcal{N}=1$ hyperbolic ( indefinite) quiver gauge
theories described by the action (\ref{act}), one rather needs the NSVZ
physical beta functions for gauge and matters\cite{31}. These are obtained
by extending affine analysis of \cite{13} to the hyperbolic (indefinite)
case.

Above $\Delta $, we get
\begin{equation}
\beta _{i}^{phys}=\sum_{j}K_{ij}^{-}N_{j}+\mathrm{\gamma }%
_{i}N_{i}+\sum_{j\neq i}\mathrm{\eta }_{ij}\left\vert K_{ij}^{-}\right\vert
N_{j};\qquad \beta _{ij}^{phys}=\beta _{ij}=\mathrm{\gamma }_{i}+2\mathrm{%
\eta }_{ij},
\end{equation}%
where $\mathrm{\gamma }_{i}=\mathrm{\gamma }\left( \Phi _{i}\right) $ and $%
\mathrm{\eta }_{ij}=\mathrm{\eta }\left( Q_{ij}\right) $ are respectively
the anomalous dimensions of massive adjoint and bi-fundamental matters and
where we have set $\beta _{i}^{phys}=\beta \left( g_{i}^{phys}\right) $ and $%
\beta _{ij}^{phys}=\beta \left( \lambda _{ij}^{phys}\right) $. The above eq
can be also rewritten into a combined form as $\beta _{i}^{phys}=\frac{%
\mathrm{\gamma }_{i}+2}{2}\left[ 2N_{i}-\sum_{j\neq i}\frac{\left( \beta
_{ij}-\mathrm{\gamma }_{i}-2\right) }{\mathrm{\gamma }_{i}+2}K_{ij}^{-}N_{j}%
\right] $.

Below $\Delta $, the $\mathrm{\gamma }_{i}$ anomalous dimensions can be set
to one, so the above beta functions reduces to $\beta _{i}=\frac{3}{2}\left(
2N_{i}+\sum_{j\neq i}\frac{\left( \beta _{ij}-\mathrm{3}\right) }{\mathrm{3}}%
\left\vert K_{ij}\right\vert N_{j}\right) $. For marginal quartic
superpotential $Tr\left[ \left( Q_{ij}Q_{ij}\right) ^{2}\right] $ and if
moreover $\beta _{ij}=0$, the anomalous dimensions $\mathrm{\eta }_{ij}$
become minus half and so the above $\mathcal{N}=1$ beta functions reduce to $%
\beta _{i}=\frac{\mathrm{\gamma }_{i}+2}{2}K_{ij}N_{j}$ which should be
compared with the value of $\mathcal{N}=2$ hyperbolic (indefinite) quiver
gauge model.

In the end of this discussion, we would like to make a note concerning
embedding of indefinite quiver gauge theories in type IIB string
compactification on CY threefolds. While hyperbolic quiver gauge models have
a meaningful type IIB string interpretation, indefinite generalizations need
however more explorations. For instance in the former subset, the gauge
coupling associated with the hyperbolic node is strongly linked with the RR
axion field, but for indefinite case the situation deserves more analysis.

\begin{acknowledgement}
This research work on brane physics enters in the framework of Protars
III/D12-25/CNR (2003) Rabat. E.H Saidi would like to thank CNR/CSIC for
support and Cesar Gomez for kind hospitality at ITF Madrid where part of
this work has been done. The work of A.Belhaj is supported by Ministerio de
Educaci\'{o}n Cultura y Deportes, (Spain) grant SB 2002-0036. R.Ahl Laamara
and L.B.Drissi are so grateful to high Energy section of ICTP for kind
hospitality.
\end{acknowledgement}

\section{Appendices}

Here we give two appendices, one regarding root system of indefinite Lie
algebras and the other deals with hyperbolic ADE quiver gauge theories.

\subsection{Appendix A: Indefinite Lie algebras}

Like for ordinary and affine Kac-Moody ADE Lie algebras, a nice way to
approach indefinite simply laced Lie algebras is in terms of simple roots $%
\alpha {_{i};}$ ${i=1,...,n}$ and corresponding Cartan matrices $K_{ij}$.
From axiomatic point of view, these require the knowledge of the triplet $(%
\mathbf{\hbar },\Pi ,\Pi ^{\nu })$ defining respectively commuting Cartan
generators ${H_{i}}$, the basis of simple roots ${\alpha }{_{i}}$ and the
basis of their duals. In representation theory; and too particularly in
quantum field theory, this triplet corresponds to specifying the commuting
observables and the basic quantum excitation operators. With these simple
roots, one may encode most of algebraic features of Lie algebras including
indefinite ones. For instance, whenever there exists a bi-linear form (,),
the Cartan matrix of simply laced indefinite Lie algebras can be realized as
usual as
\begin{equation}
K_{ii}=2;\qquad K_{ij}=2\frac{(\alpha _{i},\alpha _{j})}{(\alpha _{i},\alpha
_{i})}\in \mathbb{Z}_{-}\text{ \ for\ \ }i\neq j,  \label{11}
\end{equation}%
which is equal to $K_{ij}=\alpha _{i}.\alpha _{j}$ seen that $\mathbf{(}%
\alpha _{i},\alpha _{i}\mathbf{)}=2$. For this class of indefinite
symmetrizable simply laced Lie algebras, the statement that $K_{ij}$ is the
Cartan matrix of an indefinite Lie algebra means that for some real positive
definite vector $\mathbf{x}\in \mathbf{\hbar }$, we have,
\begin{equation}
K_{ij}x_{j}=-y_{i}<0;\qquad i,j=1,2,...,n,  \label{12}
\end{equation}%
where $x_{j}>0$, $j=1,2,...,n$ are components of $\mathbf{x}$ and $y_{i}$
are also positive, hence $\left( -y_{i}\right) $ are negative definite. The
negative sign in front of $(-y_{i})$ is remarkable since it has an
interesting geometric interpretation indicating that the space $\mathbf{%
\hbar }$ has an indefinite signature with a pseudo-euclidean metric. This
metric extends the Lorentzian signature of usual affine symmetries. Because
of this indefinite signature, indefinite Lie algebras are still a
mathematical open subject. For instance their classification has not yet
been achieved. Nevertheless, there are some subsets of these indefinite
symmetries where there are exact results. One of these subsets the \textit{%
hyperbolic }sub-class considered in this paper. According to W. Li
classification, there are $238$ hyperbolic Lie algebras containing the
following special simply laced list,

\begin{eqnarray}
&&\mathcal{H}_{1}^{4},\quad \mathcal{H}_{2}^{4},\quad \mathcal{H}%
_{3}^{4},\quad \mathcal{H}_{1}^{5},\quad \mathcal{H}_{8}^{5},\quad \mathcal{H%
}_{1}^{6},\quad \mathcal{H}_{5}^{6},\quad \mathcal{H}_{6}^{6},\quad \mathcal{%
H}_{1}^{7},  \notag \\
&&\mathcal{H}_{1}^{7},\quad \mathcal{H}_{1}^{8},\quad \mathcal{H}%
_{4}^{8},\quad \mathcal{H}_{5}^{8},\quad \mathcal{H}_{1}^{9},\quad \mathcal{H%
}_{4}^{9},\quad \mathcal{H}_{5}^{9},\quad \mathcal{H}_{1}^{10},\quad
\mathcal{H}_{4}^{10}.
\end{eqnarray}%
For instance, $\mathcal{H}_{3}^{4}$ is just the hyperbolic extension of
affine A$_{2}^{\left( 1\right) }$. More details on this sub-classification
as well as applications of hyperbolic invariance may be found in \cite{24},%
\cite{25},\cite{14,15} and refs therein.

\subsection{Appendix B: Hyperbolic ADE models}

Depending on the values of quiver gauge group ranks given by the charge of
regular D3 and fractional D5-branes, beta function components in hyperbolic
ADE models can have arbitrary signs. In general, each $\beta _{i}$ factor
can be either positive ( $\beta _{i}>0$), zero ( $\beta _{i}=0$) or negative
( $\beta _{i}<0$). This last property shows that the study of the RG flows
in hyperbolic gauge models depends on these signs. To have an idea on how to
get and fix these signs, we start from the general formula of holomorphic
beta functions in $\mathcal{N}=2$ hyperbolic gauge models,
\begin{eqnarray}
\beta _{i} &=&\sum_{j=-1}^{r}K_{ij}\widehat{N}_{j};\qquad -1\leq i\leq r,
\notag  \label{bett1} \\
\beta _{\delta } &=&\sum_{i=0}^{r}d_{i}\beta _{i}=d_{0}\beta _{0}+\beta
_{\psi },  \label{bett2} \\
\beta _{\gamma } &=&\sum_{i=-1}^{r}d_{i}\beta _{i}=d_{-1}\beta _{-1}+\beta
_{\delta },  \notag  \label{bett3}
\end{eqnarray}%
then solve them by using Lie algebra results. First note that because of the
identity $\sum_{j=0}^{r}K_{ij}d_{j}=0$ and using the fact that $K_{-1j}=0$
for $1\leq j\leq r$, the first relation of above eqs has a kernel which
leads to a remarkable simplification. Indeed taking the ranks as $\widehat{N}%
_{i}=Md_{i}$, we have
\begin{equation}
b_{i}=\sum_{j=-1}^{r}K_{ij}\widehat{N}_{j}=M\left(
K_{-1i}d_{-1}+\sum_{j=0}^{r}K_{ij}d_{j}\right) =MK_{-1i}d_{-1}
\end{equation}%
which is non zero only for $b_{-1}=2Md_{-1}$\ and $b_{0}=-Md_{-1}$. As a
consequence of this feature, the beta functions $\beta _{i}$, $1\leq i\leq r$
of hyperbolic ADE quiver gauge theories are independent on the number of
regular branes; they behave exactly as in ordinary ADE quiver gauge models.
So, the quantity $\sum_{j=-1}^{r}K_{ij}\widehat{N}_{j}$ reduces to
\begin{eqnarray}
\beta _{-1} &=&N_{-1}-N_{0};\qquad \beta _{0}=-\left(
N_{-1}+\sum_{j=1}^{r}\left\vert K_{0j}\right\vert N_{j}\right) <0, \\
\beta _{i} &=&\sum_{j=1}^{r}K_{ij}N_{j};\qquad 1\leq i\leq r,  \notag
\end{eqnarray}%
showing that the previous results gotten for hyperbolic A$_{r}$ models are
also valid for hyperbolic DE ones. Therefore in hyperbolic\ ADE\ gauge
theories, $\beta _{0}$ is usually negative and the sign of $\beta _{-1}$ is
given by the signature of $\left( N_{-1}-N_{0}\right) $. To get the sign of
remaining $\beta _{i}$; we consider the UV regime where $\mu >\Lambda _{i}$
with $1\leq i\leq r$. This domain can be represented in a simpler way by
ordering gauge scales $\Lambda _{i}$ as $\Lambda _{1}>\Lambda
_{2}>...>\Lambda _{r}$. In this parameterizations, positivity condition of
beta components translates into $\mu >\Lambda _{1}$. This domain has a nice
interpretation in Lie algebra language. Precisely it corresponds to the
Vinberg phase which states that there usually exists a configuration
\begin{equation}
\mathbf{N}=\left( N_{1}^{\left( 0\right) },...,N_{r}^{\left( 0\right)
}\right) ;\qquad N_{i}>0,  \label{np}
\end{equation}%
such that $\beta _{i}>0$ with $1\leq i\leq r$. So UV regime in ordinary
supersymmetric quiver gauge theories corresponds to Vinberg configuration in
finite dimensional Lie algebras. As Vinberg theorem is in connection with RG
flows in supersymmetric quiver gauge theories, we will give details on this
issue. The system (\ref{bett2}) defining $\beta _{i}$s is very impressive;
it recalls the Vinberg theorem on Lie algebras classification. This theorem
states that for any Lie algebra, there exists two \textit{positive real}
vector $\mathbf{u}=\left( u_{1},...,u_{n}\right) $ and $\mathbf{v}=\left(
v_{1},...,v_{n}\right) $ such that,
\begin{equation}
\sum_{\rho }K_{\mu \rho }^{\left( q\right) }u_{\rho }=qv_{\rho }  \label{bit}
\end{equation}%
where $K_{\mu \rho }^{\left( q\right) }$ with $q=+1,0,-1$ stand respectively
for the Cartan matrices of ordinary, affine and hyperbolic ( indefinite) Lie
algebras. In our case, the correspondence is given by $\mathbf{%
u\longleftrightarrow N}=\left( N_{-1},...,N_{r}\right) $ and $\mathbf{%
v\longleftrightarrow \beta }=\left( \beta _{-1},...,\beta _{r}\right) $.
Following this theorem and using known results on geometric engineering on $%
\mathcal{N}=2$ supersymmetric QFT$_{4}$s \cite{2,14} stating that
holomorphic beta functions are proportional to Lie algebra Cartan matrices,
one\ sees that in general supersymmetric quiver gauge theories have three
remarkable phases parameterized by the quantum number $q=-1,0,-1$. This
quantum number captures the global signs of the beta function vector $%
\mathbf{\beta }=\left( \beta _{-1},...,\beta _{r}\right) $ and so we have
the following correspondence:%
\begin{eqnarray}
q &=&+1\qquad \Longleftrightarrow \qquad \text{UV free }\mathcal{N}=2\text{
QFT}_{4}  \notag \\
q &=&0\qquad \ \ \Longleftrightarrow \qquad \text{Conformal }\mathcal{N}=2%
\text{ QFT}_{4} \\
q &=&-1\qquad \Longleftrightarrow \qquad \text{Confined }\mathcal{N}=2\text{
QFT}_{4}.  \notag
\end{eqnarray}%
According to $\beta _{i}$ signs (i.e $\beta _{i}>0$, $\beta _{i}=0$, $\beta
_{i}<0$ ), there are $3^{n}$ others more.

\end{document}